\documentclass[12pt,aps,prb,preprint]{revtex4}   

\usepackage{amsmath}    
\usepackage{amssymb}
\usepackage{graphicx}   

\usepackage[latin1]{inputenc}     
\usepackage[T1]{fontenc}

\usepackage{float}   

\draft

\begin{document}
\title{Proposed experiment with Rydberg atoms to test the wave function interpretation}
\author{Michel Gondran}
 \affiliation{EDF, Research and Development, 92140 Clamart, France}
 \email{michel.gondran@chello.fr}   
 \author{Mirjana Bo$\check{z}$ic}
 \affiliation{Institut of Physics, P.O. Box 57, 11001 Belgrade, Serbia and Montenegro, Yugoslavia}
 \email{bozic@phys.bg.ac.yu, arsenovic@phys.bg.ac.yu}
\author{Alexandre Gondran}
 \affiliation{SET, Université Technologique de Belfort Montbéliard, France}
 \email{alexandre.gondran@utbm.fr}   
\author{Du$\check{s}$an Arsenovi$\acute{c}$}
 \affiliation{Institut of Physics, P.O. Box 57, 11001 Belgrade, Serbia and Montenegro, Yugoslavia}
 \email{arsenovic@phys.bg.ac.yu}   

\begin{abstract}

Experiment~\cite{Fabre_1983} shows that Rydberg atoms do not pass
through $1 \mu$ m width slits if their principal quantum number is
rather large($n\geq 60$). Thus, the particle density measured
after the slits is null while the wave function calculated after
the slits is not. This experiment is in contradiction with the
Born interpretation (the square of the wave function is
proportional to the probability density for the particle to be
found at each point in space). The classical interpretation of
this experiment, which removes the contradiction, is to suppose
that if the particles do not pass, the wave function does not pass
either (\textit{classical assumption}).

An alternative interpretation of this experiment is to suppose
that the wave function passes through the slits, but that the Born
interpretation is not valid any more in this case
(\textit{alternative assumption}).

The aim of this paper is to present an experiment testing this
\textit{alternative assumption} compared to the \textit{classical
assumption}.

\end{abstract}

 \maketitle

\section{Introduction} \label{}

\bigskip

One of the fundamental postulates of quantum mechanics is the Born
interpretation, where the square of the module of the wave
function is equal to the particle density. However, for a long
time, there were no experiments that could separate the particle
from its wave function; the wave function accounts for the mass,
the wavelength, the spin and the particle electric charge. On the
other hand, the wave function does not account for the particle
size. However, a quite old experiment of Haroche's
team~\cite{Fabre_1983} accounts explicitly for Rydberg atoms size.
It is an experiment where one measures "\textit{the transmission
of a beam of Rydberg atoms through a metallic grating made of an
array of micrometre size slits. The transmission decreases
linearly with the square of the principal quantum number n, with a
cut-off for a maximum n value}". This experiment inspired recently
Dahl \textit{et al.}~\cite{Dahl_2005} to study theoretically the
scattering of a rotor from a slit in the limit when the length of
the rotor is larger than the size of the slit.

Thus, as soon as n is rather large ($n\geq 60$ in this experiment
for $1 \mu m$ width slits), the particle density measured after
the slits is null while the wave function calculated after the
slits is not. This is in contradiction to the Born interpretation.

The usual interpretation of this experiment, which removes the
contradiction, is to suppose that if the particles do not pass,
the wave function does not pass either (\textit{classical
assumption}). This interpretation, which seems obvious for the
majority of the quantum mechanics community, was however never
tested in experiments; the presence of the wave function being
measured only via the impact of  particles.

Although not very probable, one cannot thus eliminate logically
alternative interpretation of this experiment which supposes that
the wave function passes through the slits but not the particles
(\textit{alternative assumption}). In this case, the Born
interpretation is not valid.

The aim of this paper is to present an experiment testing this
\textit{alternative assumption} compared to the \textit{classical
assumption}.

The idea of this experiment is initially based on the fact that
diffraction and interference phenomena exist for large particles,
cf. Schmiedmayer \textit{et al}~\cite{Schmiedmayer_1993} and
Chapman \textit{et al}~\cite{Chapman_1995} for molecules of $Na_2$
($\sim 0.6 nm$ size) and the Zeilinger team for the fullerene
molecules $C_{60}$ ($\sim$ 1 nm diameter)~\cite{Zeilinger_1999},
$C_{70}$~\cite{Zeilinger_2000} and more recently the
fluorofullerenes $C_{60}F_{48}$~\cite{Zeilinger_2003}. It is also
based on Bozic \textit{et al}~\cite{Bozic_2002,Bozic_2004} thought
experiments which consider interference with different size slits.

Section 2 describes this test experiment which is an interference
experiment with Rydberg atoms and asymmetrical slits; one wide
slit A let the Rydberg atoms pass, and one grid B of small slits
does not let them pass.

Section 3 gives the value of the square wave function after the
slits in the case of both assumptions, the \textit{classical
assumption} and the \textit{alternative assumption}.

Finally, in section 4, we present the atomic density which we
should measure on the detection screen in the case of the
alternative hypothesis. One also supposes in
 this case that the Rydberg atoms pass through the slit A but do not pass through the grid B.

\section{The Rydberg atoms asymmetric experiment}

An asymmetric Rydberg atoms experiment, inspired by the Fabre
\textit{et al} experiment~\cite{Fabre_1983}, is proposed. A
Rydberg atoms' beam with a principal quantum number $n =60$, whose
speed is $v_y = 200 m/s$ along the 0y axis, is considered. Initial
speeds in the other directions are considered null.


The beam is regarded as Gaussian with 6mm width.

At the distance $d_1=1$m from the molecular beam source, a metal
foil is placed that has a slit A with a 100$\mu$m slit width along
the axis 0x and a grating B with 1000 small slits of 0.1 $\mu$m
slit width and 0.2$\mu$m slit separation along the 0x axis. The
distance between the centers of slit A and grating B is 300$\mu$m.
Rydberg atoms are then observed by using a detector placed at
$d_2=2$m behind the slits. The Rydberg atom beam is obtained from
sodium atoms, so the atom mass is $m=3.84$ $10^{-26}$kg. With the
velocity $v_y = 200 m/s$, the de Broglie wavelength
$\lambda_{dB}=8.6 \times 10^{-5}$$\mu$m is much smaller than the
slits size.

The slit size of the grid B (0.1 $\mu$m) is ten times smaller than
for the Fabre \textit{et al} experiment~\cite{Fabre_1983}. In this
case, the Rydberg atoms with n=60 do not pass through the grid. On
the other hand, slit A is 100 times larger than the Rydberg atoms'
size and thus lets them pass easily.

For n=60, the 70 ms of the Rydberg atom lifetime are sufficient
for the experiment (15 ms).

\begin{figure}[H]
\begin{center}
\includegraphics[width=0.6\linewidth]{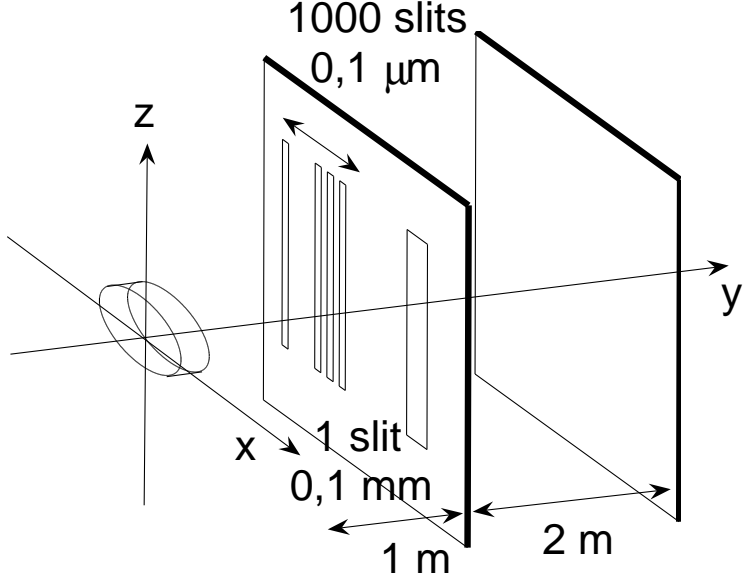}
\caption{\label{fig:schemexperience} Schematic diagram of the
Rydberg atoms asymmetric experiment.}
\end{center}
\end{figure}

The case of the classical assumption where the wave function does
not pass through the grid B corresponds to diffraction by the slit
A.

The case of the alternative assumption where the wave function
passes through the grid B corresponds to an interference problem
between the slit A and the grid B slits.

\section{Wave function calculation with Feynman path integral}

The wave function calculation is obtained by a numerical
calculation using Feynman integrals, as we did~\cite{Gondran_2005}
for the numerical simulation of Shimizu \textit{et al }'s
experiment with cold atoms.

The wave function before the slits is then equal to
\begin{equation}\label{eq:eqavantfentes}
	\psi(x,t)=(2\pi s(t)^2)^{-\frac{1}{4}}\exp^{-\frac{x^2}{4
\sigma_0 s(t)}}
\end{equation}
with $s(t)=\sigma_0 (1+\frac{i \hbar t}{2 m \sigma_0^2})$. After
the slits, at time $t\geq t_1=\frac{d_1}{v_y}=5~ms$, the
time-dependent wave function $\psi(x,t)=\psi_A(x,t) + \psi_B(x,t)$
is calculated by the Feynman path integral
method~\cite{FeynmanQMI}:
\begin{equation}\label{eq:eqapresfentes}
	\psi_{A / B}(x,t)=\int_{A / B} K(x,t;x_f,t_1) \psi(x_f,t_1) dx_f
\end{equation}
where $\psi(x_f,t_1)$ is given by (\ref{eq:eqavantfentes}) and
\begin{equation}\label{eq:eqdeK}
	K(x,t;x_f,t_1)=\left(\frac{m}{2 i \pi
\hbar(t-t_1)}\right)^{\frac{1}{2}}\exp\frac{im}{\hbar}\frac{(x-x_f)^2}{2(t-t_1)}.
\end{equation}
The integrations in (\ref{eq:eqapresfentes}) are carried out
respectively on the slit A and on the grid B slits.

\begin{figure}[H]
\begin{center}
\includegraphics[width=0.45\linewidth]{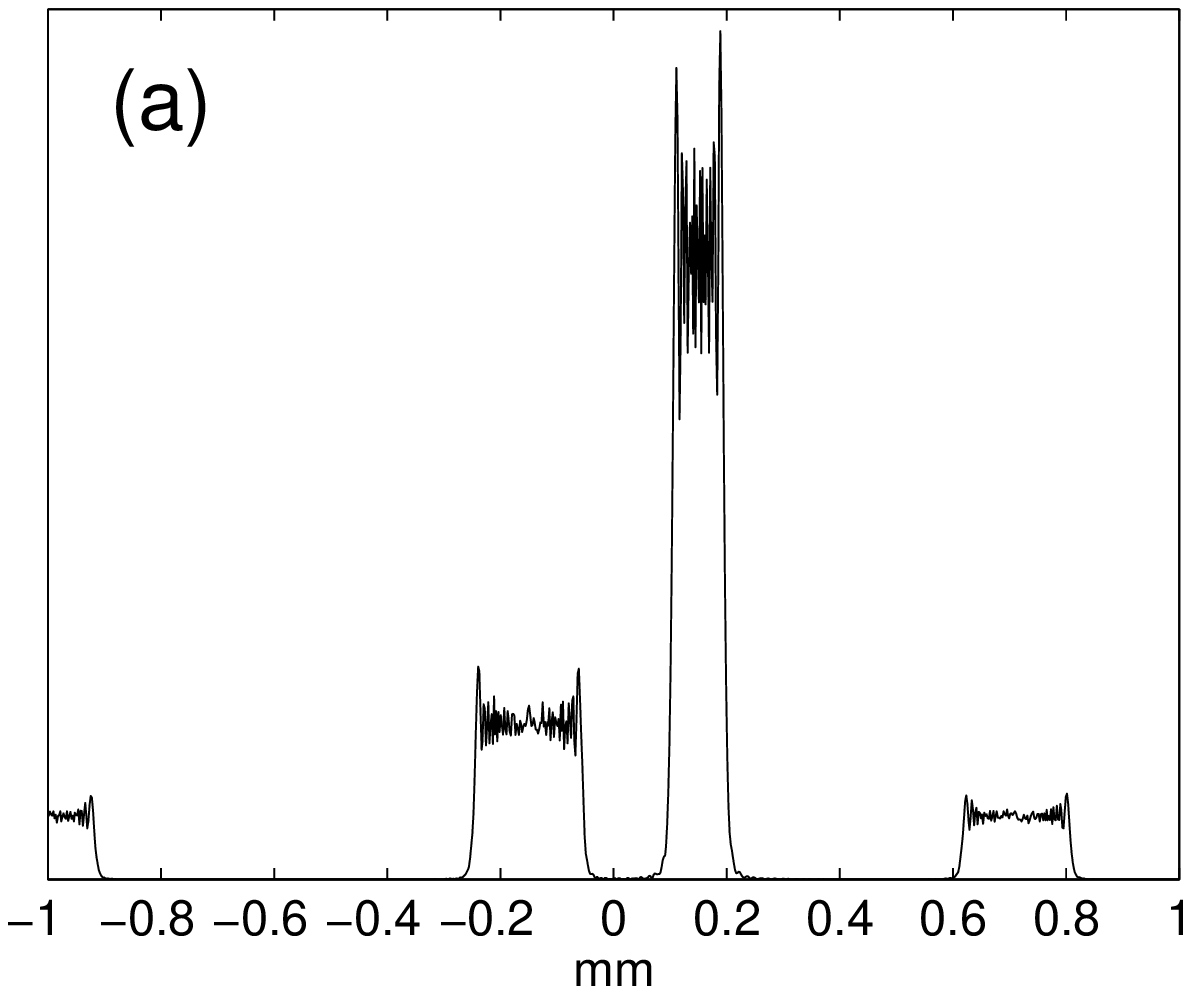}
\includegraphics[width=0.45\linewidth]{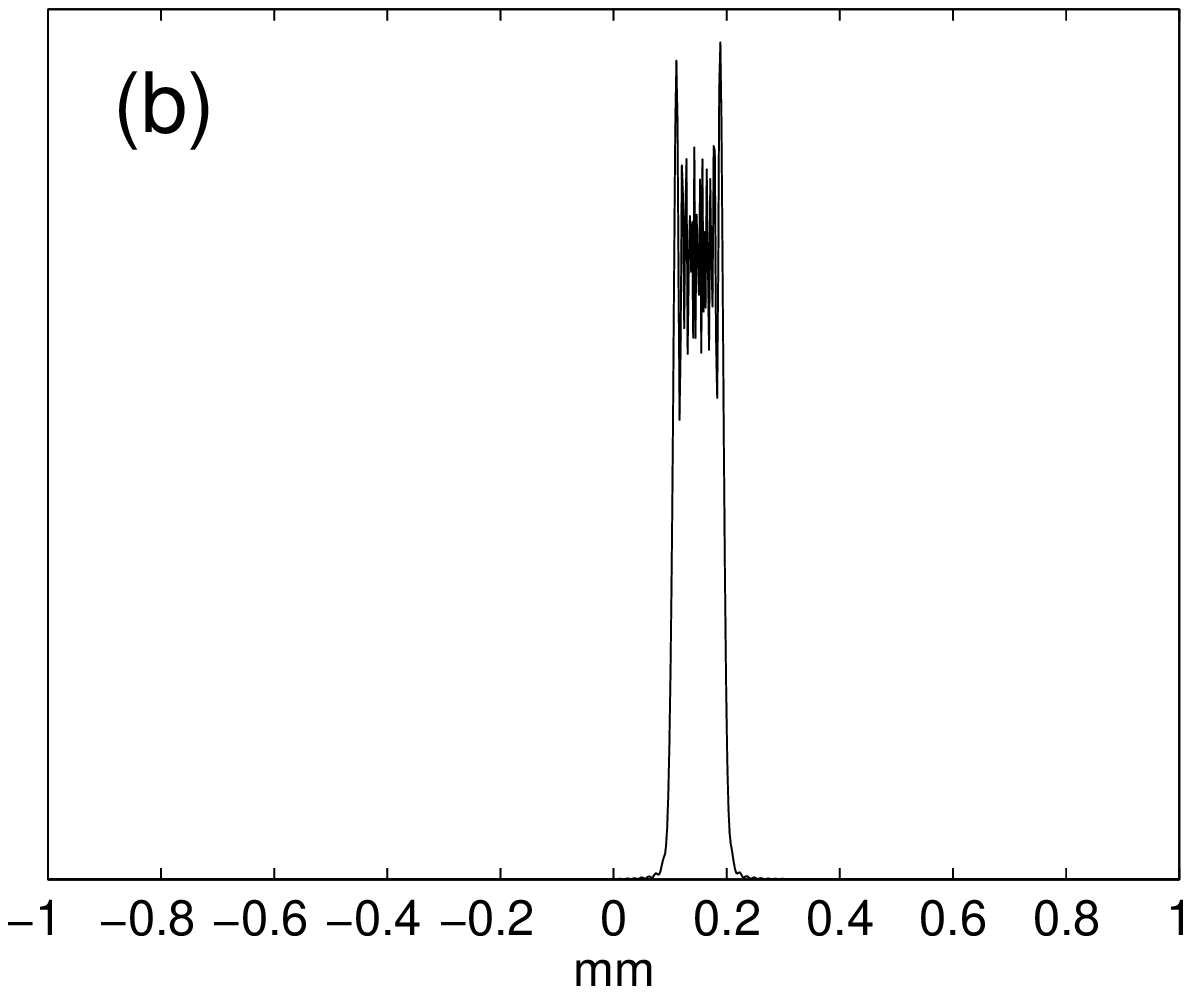}
\caption{\label{fig:comparaisonInterferenceDiffractions} Square of
the wave function $|\psi(x)|²$ at 2 m from the slits (a) in the
case of interference between A and B, and (b) in the case of
diffraction by A.}
\end{center}
\end{figure}

Figure 2 represents the square of the wave  function
$|\psi(x,t)|²$ on the detector at 2 m behind the slits for the
interference experiment between the slit A and the grating B
(figure 2a), and for the diffraction experiment with the slit A
only (figure 2b). For the diffraction case, we obtain classically
one peak, and for the interference case, we obtain distinctly four
peaks.

Figure 3 details the $|\psi(x,t)|²$ evolution according to the
distance to the slits in the case of interference between A and B.
\begin{figure}[H]
\begin{center}
\includegraphics[width=0.60\linewidth]{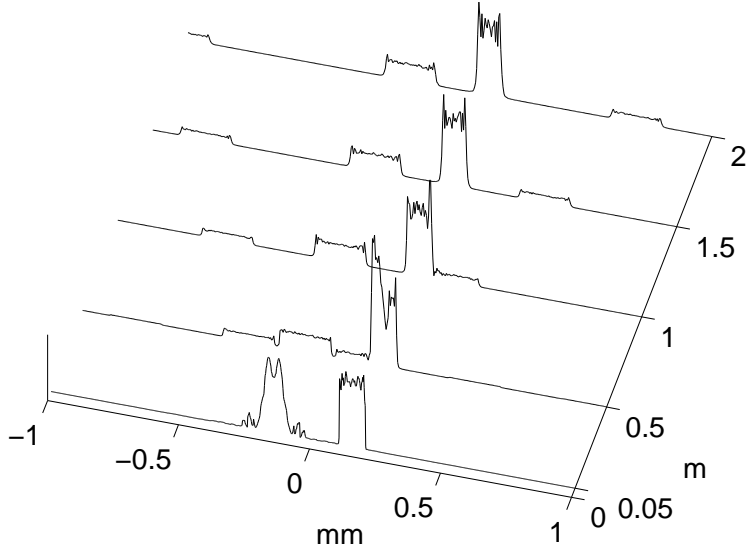}
\caption{\label{fig:4} Evolution of the square wave function
$|\psi(x,t)|²$ in the first 200 cm behind the slits for the case
of interference between A and B. Note that $\psi(x,t)$ is
equivalent to $\psi(x,y)$ because $t=y/v_y + t_1$ with
$v_y=200m/s$. So calculating the wave function at 2m from the
slits is equivalent to calculating the wave function 10ms after it
passes through the slits ($y/v_y=2/200=0.01s$)}
\end{center}
\end{figure}

If the particles size is not considered (\textit{usual
assumption}), the wave function and the particles would "pass
through" the slit A \textbf{and} the grid B. The particle density
measured on the detector will then have to correspond to the four
peaks of figure 2a (interferences phenomenon between A and B).

In the case of the \textit{classical assumption}, the wave
function and the particles "pass through" the slit A \textbf{but
"do not pass through"} the grid B. The particle density measured
on the detector will have to correspond to the single peak of
figure 2b (diffraction phenomenon by slit A).

We study in the next paragraph the \textit{alternative assumption}
where:
\begin{itemize}
	\item the wave function "passes through" the slit A \textbf{and}
the grid B, 	\item the particles "pass through" the slit A
\textbf{but "do not pass through"} the slits of grid B.
\end{itemize}

\section{Alternative assumption}

In the traditional experiments of double slits carried out with
molecules whose size is smaller than the slits, the wave function
passes through the two slits to form interferences, but each
molecule passes through one or the other of the slits, without
knowing through which slit it passed. The idea of the alternative
assumption is the same. The wave function passes through slit A
and grid B and thus will create the interferences of figure 2a.

However, the atoms do not pass through the slits of the grid B.
One deduces from this that, just behind the grid B, the
probability density of the atoms must be null since no atom
crossed it. In the case of the \textit{alternative assumption},
$|\psi(x,t)|²$ thus no longer represents the probability density
of the atoms.

We are in a nonusual situation in quantum mechanics. However, as
we already showed in~\cite{Gondran_2005}, we verify that, just
after the slits at time $t_1^+$, there is not yet covering between
the wave function $\psi_A$ which passed by A and the wave function
$\psi_B$ which passed by B and we have $|\psi_A(x,t_1^+)+
\psi_B(x,t_1^+)|^2=|\psi_A(x,t_1^+)|^2+|\psi_B(x,t_1^+)|^2$.

Since the atoms do not pass by the grid B, one deduces from it
that the density $\rho((x,t_1^+)$ of the atoms which passed is
equal to $|\psi_A(x,t_1^+)|^2$, the part of the right-hand side of
$|\psi(x,t_1^+)|²$, cf. figure 3 at time $t_1 + 0.25$ ms (0.05 m).

Since the sum of the widths of the slits of the grid B is equal to
the width of slit A, the density $\rho((x,t_1^+)$ is thus
\begin{equation}\label{eq:eqdensitet1}
\rho(x,t_1^+)=
\begin{cases}
    0 & \text{if}~~~~-\infty < x < x_ {t_1^+}, \\
    |\psi(x,t_1^+)|² & \text{if}~~~~x_ {t_1^+}< x < +\infty
  \end{cases}
\end{equation}
where $x_ {t_1^+}$ is the median of the function
$F(x,t)=\int_{-\infty}^x
\frac{|\psi(x,t)|²}{\int_{-\infty}^{+\infty}|\psi(u,t)|²du} du$,
i.e. such as $F(x_{t_1^+},t_1^+)=\frac{1}{2}$. By continuity, we
can suppose that the atoms, which passed through the slit A,
satisfy at each time this property: the density is thus
\begin{equation}\label{eq:eqdensitet2}
\rho(x,t)=\begin{cases}
    0 & \text{if}~~~~-\infty < x < x_ {t}, \\
    |\psi(x,t)|² & \text{if}~~~~x_ {t}< x < +\infty
  \end{cases}
\end{equation}
where $x_t$ is the median of the function $F(x,t)$.

Figure 4 details the density evolution of the atoms having passed
throug slit A according to the distance to the slits.

\begin{figure}[H]
\begin{center}
\includegraphics[width=0.60\linewidth]{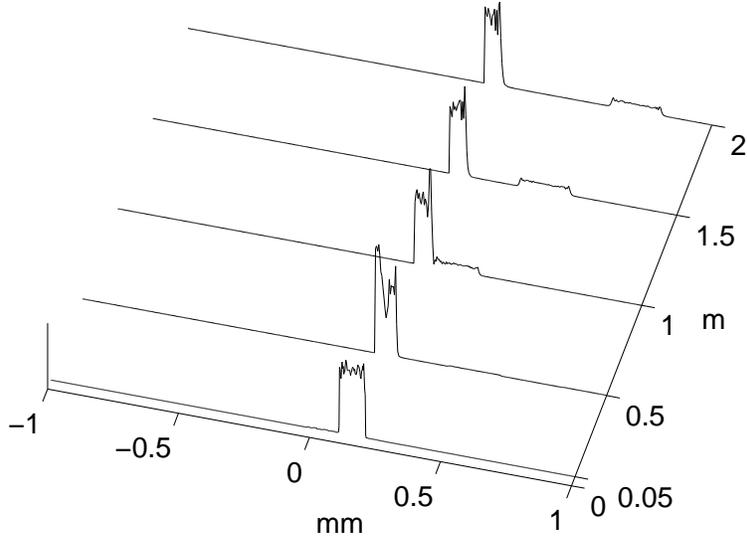}
\caption{\label{fig:5} Evolution on the atoms density in the first
200 cm after slits for the case of the \textit{alternative
assumption}.}
\end{center}
\end{figure}

Consequently, at the time $t_2= 15$ms, i.e. on the detector, the
particle density will be equal to the density located on the right
of the median. Figure 5 shows this density corresponding to the
right-hand side of figure 2a.

\begin{figure}[H]
\begin{center}
\includegraphics[width=0.5\linewidth]{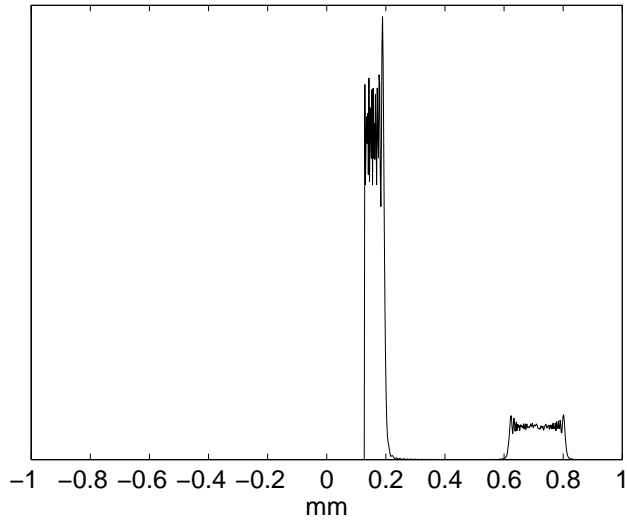}
\caption{\label{fig:6} Particle density at 2 m from the slits in
the case of the \textit{alternative assumption}.}
\end{center}
\end{figure}

Another way of taking into account the alternative assumption
assumes the existence of the trajectories. The trajectories may be
de Broglie-Bohm trajectories~\cite{Bohm, Holland, Holland_1998,
Gondran_2005,Gondran_2006} or the trajectories approximately
determined by transverse momenta of particles and their
distribution~\cite{Bozic_2002b, Bozic_2004}. In this case, the
particles which follow the trajectories and have to pass through
the grid B are stopped. The particles having to pass through slit
A pass; and their trajectories correspond to the evolution of the
densities of figure 4. One thus obtains the same result as figure
5, but with an assumption unnecessary.

The clear difference between the density of figures 2b and 5 shows
that the asymetric Rydberg atoms experiment can be a very good
test between the \textit{classical assumption} and the
\textit{alternative assumption}. This test is robust in spite of
the variation of the initial data with the initial speeds $v_x$,
$v_y$ and $v_z$. We have shown \cite{Gondran_2005} how to take
these uncertainties into account. They will smooth the densities,
but will preserve the number of peaks.

\bigskip

\section{Conclusion}

Taking into account the size of the Rydberg atoms in the
interference phenomena makes it possible to renew the study of the
wave-particle duality and to propose an experiment to test the
wave function interpretation.

One tests the traditional interpretation that if the particles do
not pass through a slit, the wave function does not pass either,
against the \textit{alternative assumption} that the wave function
passes through the slit but that the interpretation of Born is not
valid any more in this case.

This experiment appears realizable today, and each of the two
alternatives will contribute an interesting contribution to the
interpretation of the wave function. Indeed, even in the case
considered as most probable where the experiment confirms  the
\textit{classical assumption}, this result will have to be to add
to the other postulates of quantum mechanics.

\end{document}